\documentstyle[12pt]{zer}
\newcommand{\non}{\nonumber}

\begin{document}
\begin{titlepage}

\begin{flushright}
DESY 94--076 \\
BUDKERINP--94--39 \\
hep-ph/9405301\\
April 1994
\end{flushright}

\vspace{0.3cm}

\begin{center}
{\Large \sc Threshold Effects in Two--Photon Decays \\[0.5cm]
of Higgs Particles}
\end{center}

\vspace{0.8cm}

\begin{center}
{\large Kirill Melnikov$^1$, Michael Spira$^2$ and Oleg Yakovlev$^{2,3}$}

\vspace{0.8cm}

\noindent
     $^1$ Novosibirsk State University, Novosibirsk--90, Russia \\
     $^2$ Theory Division, DESY, D--22603 Hamburg, FRG \\
     $^3$ Budker Institute for Nuclear Physics, Novosibirsk--90, Russia

\end{center}

\vspace{0.8cm}

\begin{abstract}
The couplings of Higgs particles to two photons are analyzed in the
threshold region, where the Higgs mass is about twice the loop--particle mass
leading to possible bound state formation. The amplitude of the pseudoscalar
decay $A\to \gamma\gamma$ for pseudoscalar masses in the range of
$(t\bar t)$--bound state masses is
calculated in leading and next--to--leading order in $\alpha_s$.
It is shown that the real and imaginary parts of the form factor are
significantly different from the lowest order perturbative one.
\end{abstract}

\end{titlepage}

\section{Introduction}
%        ============

The Standard Model [${\cal SM}$] has been tested to a very high precision in
recent years. However, the Higgs boson, one of the elementary particles of this
model, has not been discovered until now.
Higgs boson hunting remains one of the most important problems
in contemporary high energy physics.
Recently the two--photon decay mode of the Higgs boson attracted much
attention by both theorists and experimentalists
based on two different reasons:
\begin{enumerate}
\item This decay channel provides an opportunity to discover and
study the Higgs boson in the lower part of the intermediate mass region
at hadron colliders such as the LHC \cite{Kunz}.

\item Two--photon production of a heavy Higgs boson via
$\gamma \gamma \to H\to ZZ, b\bar b, t\bar t$ is quite interesting
and promising for future high energy $\gamma\gamma$ colliders \cite{Zerw}.
Scalar Higgs production via photon fusion exhibits a possibility to distinguish
between the Higgs boson of the ${\cal SM}$ and the
light scalar one of the minimal supersymmetric extension of the
Standard Model [${\cal MSSM}$] in the case of the pseudoscalar
being very heavy and to measure the Higgs parity \cite{Kraem}.
\end{enumerate}

In this paper we consider the coupling of scalar as well as pseudoscalar
Higgs bosons to two photons. The existence of
pseudoscalar Higgs bosons is predicted by
e.g.~the ${\cal MSSM}$. This model contains
three neutral Higgs particles: two $\cal CP$--even scalars $h,H$
and one $\cal CP$--odd pseudoscalar $A$.
The $\gamma \gamma$ couplings to these particles
are mediated by charged heavy particle
loops ($W$--boson, $b$-- and $t$--quark) as shown in fig.1.
 The top quark and $W$ amplitudes read in lowest order \citer{Elli,Gun}
\begin{eqnarray}
F_t^H & = & 2 \tau [1+(1-\tau) f(\tau)] \non \\ \non \\
F_W^H & = & - [3\tau+2 -3\tau (\tau-2) f(\tau)] \non \\ \non \\
F_t^A & = & \tau f(\tau) \label{eq:ff}
\end{eqnarray}
The scaling variable is defined as $\tau =4m_i^2/m_\phi^2$,
where $m_i$ denotes the loop--particle mass and
$m_\phi$ the corresponding Higgs mass, and
\begin{equation}
f(\tau) = \left\{ \begin{array}{ll}
\displaystyle
\arcsin^2 \frac{1}{\sqrt{\tau}} & \tau\geq 1 \\ \\
\displaystyle
-\frac{1}{4} \left( \log\frac{1+\sqrt{1-\tau}}
{1-\sqrt{1-\tau}} -i\pi \right)^2 & \tau < 1
\end{array} \right.
\end{equation}
The two--loop QCD radiative corrections to the quark amplitudes have been
calculated for the scalar Higgs decays \citer{Zhen,MeYa} and the pseudoscalar
one \cite{DSZ}.
Near the $t\bar t$--threshold [$\tau\sim 1$] the perturbative result of these
corrections to the pseudoscalar top--loop is not valid due to
the appearance of a Coulomb singularity originating from possible
$(t\bar t)$--bound state formation in an $S$--wave. In this mass region the
narrow width approximation is not applicable.
Due to the quantum numbers of the Higgs particles the intermediate
$t\bar t$--pairs produced in the scalar Higgs decays $h,H \rightarrow t\bar t
\rightarrow \gamma\gamma$ belong to a $P$--wave amplitude, which is strongly
suppressed at threshold leading to negligible effects of the top decay width
and Coulomb interactions of the $t\bar t$--pair. Therefore we only consider
effects of the $W$ decay width in the scalar decays to photons.
The perturbative results for the lowest order amplitudes
(\ref{eq:ff})
and the two--loop QCD corrections at threshold are significantly modified
due to the following effects:
\begin{description}
\item[(i)]  strong Coulomb interaction of the loop--particles,
\item[(ii)] the large width of the $t$--quark and $W$--boson.
\item[(iii)] mixing of the Higgs bosons with the $t\bar t$--bound states
\cite{DreHi}.
\end{description}

Calculating the form factor of the process $A\to \gamma\gamma$ for
$|m_A-2m_t|\ll m_A$ we have to take into account Coulomb interactions
between $t$ and $\bar t$.
The physical picture is obvious: above threshold, the Coulomb interaction
leads to a non--zero probability of $t\bar t$ production at $\tau \to 1$,
i.e.~the imaginary part of $F^A_t$ takes the form \cite{Somm,Sakh}
\begin{eqnarray}
\Im m F_t^A & = & \Im m F_t^{A (LO)} C_{Coul} \non \\ \non \\
C_{Coul} & = & \frac{C_F \frac{2\pi\alpha_s}{v}}
{1-\exp\left(-C_F \frac{2\pi\alpha_s}{v}\right) }
\end{eqnarray}
The parameter
$v=\sqrt{1-\tau}$ is the relative velocity of the $t\bar t$ pair and
$C_F=4/3$ denotes the usual SU(3)--color factor.
The process $A\to t\bar t \rightarrow \gamma\gamma$ proceeds via an
$S$--wave resulting in an
imaginary part of
$F^{A}_t$ in lowest order proportional to $v$ at threshold.
After including the perturbative QCD corrections in the narrow width
approximation the imaginary part of $F_t^A$ develops a finite step at the
$t\bar t$--threshold, whereas the real part acquires the corresponding
logarithmic singularity for $m_A = 2m_t$. This singular behaviour of
the form factor is removed by including the finite decay width of
the top quark and Coulomb interactions leading to resonance formation
of bound states with quantum numbers $J^{\cal PC} = 0^{-+}$ below threshold.
 We should note that non--perturbative QCD effects are very important
for the bound--state formation, but due to the large width of
the top quark they are suppressed in the production
of a $t\bar t$ system [the relative correction being less than 1\%]
\citer{FaKh,Pesk}.
In this paper we take into account the finite width $\Gamma_t$
of the top quark and Coulomb interaction effects in the threshold region.
Qualitatively this analysis leads to the appearance of a finite tail
in the imaginary part of $F^A_t$
below threshold and resonance formation of
bound states with a finite width $2\Gamma_t$.
We present the results for the amplitudes $H\to \gamma\gamma$
and $A\to \gamma\gamma$ in leading and next--to--leading order in $\alpha_s$.

In order to use those amplitudes for the prediction of the Higgs production
cross-section in the vicinity of the loop-particle threshold, one has to
include mixing of the Higgs field with $t\bar t$--bound states, resulting
in important contributions to the decay width in the threshold region. An
extensive study of these effects has been performed in Ref.\cite{DreHi},
where reader can find all relevant formulas. In the following we will neglect
these mixing effects, so that our results correspond to the decays of unmixed
Higgs bosons.

\section{Coulomb Interactions}
%        ====================

The general formula for the amplitudes $F^A_t$ and $F^H_W$
for vanishing decay widths of the top quark and $W$--boson in the threshold
region reads as [see Appendix]:
\begin{equation}
F = A + B~G(0,0;E) \label{eq:green}
\end{equation}
$A$ and $B$ are real constants, determined by a non--relativistic expansion
around the threshold of the lowest order form factors (\ref{eq:ff}).
$G(0,0;E)$ denotes the $S$--wave Green's function of the non--relativistic
Schr\"odinger equation with a potential $V(r)$:
\begin{equation}
\left[ -\frac{\nabla^2}{m} + V(r) - E \right] G(\vec{r},\vec{r}\,';E) =
\delta (\vec{r} -\vec{r}\,') \label{eq:schr}
\end{equation}
First we consider the pseudoscalar amplitude $F_t^A$. To determine the
Green's function $G_t(0,0;E)$ we have to solve the Schr\"odinger equation
(\ref{eq:schr}) with $m=m_t$ and the Coulomb potential $V(r)$:
\begin{equation}
V(r) = - \frac{4}{3} \frac{\alpha_s}{r}
\end{equation}
In this section we choose the strong coupling $\alpha_s$ to be constant,
whereas effects of a running coupling $\alpha_s(r)$ will be discussed in
the next section. Non--perturbative effects of the potential $V(r)$ leading
to confinement at large distances are suppressed by the large top decay
width and will be neglected in our analysis.

Equation (\ref{eq:green}) has a clear physical meaning for this amplitude:
the process $A \to t\bar t \to \gamma \gamma $ is mediated by a
quark loop, which quarks contain a large virtuality, or by
the formation of $(t\bar t)$--bound states. These are produced near the origin
$r=0+{\cal O}(m_t^{-1})$, evolve during a time
$\tau \sim |E|^{-1}$ and annihilate to two photons at
the same point $r=0+{\cal O}(m_t^{-1})$.  So the first term in
eq.(\ref{eq:green}) has a relativistic origin from virtual quark
contributions, but the second
one is connected with the non--relativistic dynamics of the $t\bar t$ system.
Due to bound state formation the form factor $F(E)$ develops poles
at $E=E_n$. The constants $A$, $B$ as well as the position of these poles
and their residues can be expressed as expansions in $\alpha_s/\pi$:
\begin{equation}
A = \sum_n A_n \left(\frac{\alpha_s}{\pi}\right)^n \hspace{2cm}
B = \sum_n B_n \left(\frac{\alpha_s}{\pi}\right)^n \label{eq:anbn}
\end{equation}
Our main goal is the calculation of the short distance coefficients
$A, B$ in leading and next--to--leading order in $\alpha_s$, i.e.~the
determination of $A_0$, $B_0$, $A_1$ and $B_1$.
In case of a fixed value of $\alpha_s$ these can be obtained from the
well--known expression of the
Green's function calculated in Ref.\cite{Volo,FaKh,Meix}:
\begin{equation}
G_t(0,0;E) = -\frac{m_t p}{4\pi} + \frac{m_t p_0}{2\pi} \log\left(
\frac{m_t}{p} D \right) + \frac{m_t p_0^2}{2\pi} \sum_{n=1}^\infty
\frac{1}{n(np-p_0)} \label{eq:meix}
\end{equation}
with $p_0=\frac{2}{3}m_t\alpha_s$ and $p=\sqrt{m_t(-E-i\epsilon)}$.
$D$ is a real constant depending on the renormalization of the Green's
function.
The first term in eq.(\ref{eq:meix}) corresponds to the lowest order diagram,
the second to the
contribution of one coulombic gluon exchange, whereas the
third one describes the sum over all bound states with level number $n$
corresponding to diagrams with exchanges of $n$ Coulomb gluons.
We note that the uncertaincy from choosing $D$ in (\ref{eq:meix}) is of
next--to--leading order in $\alpha_s$, so that
we have to calculate the amplitude $F$
in the two--loop approximation in order to define this constant.
These QCD radiative corrections will be discussed in the next section.
We choose $D=1$ in this section for the numerical analysis.

The Green's function of the $W$--loop contributing to the scalar decays
$h,H \to \gamma\gamma$ is given by the first term of eq.(\ref{eq:meix})
being related to the solution of the Schr\"odinger equation (\ref{eq:schr})
without the potential $V(r)$, because $W$--bosons do not interact strongly
in ${\cal O}(\alpha_s)$:
\begin{equation}
G_W(0,0;E) = -\frac{m_W p}{4\pi}
\end{equation}
with $p=\sqrt{m_W(-E-i\epsilon)}$.

In order to fix the constants $A$ and $B$ we start with the
form factors $F$ in one--loop approximation (\ref{eq:ff}).
We obtain from a non--relativistic expansion at the threshold:
\begin{equation}
\begin{array}{llllll}
A_W^H & = & \displaystyle 5+\frac{3}{4} \pi^2 \hspace{2cm}
& B_W^H & = & \displaystyle - \frac{12 \pi^2}{m_W^2} \\ \\
A_t^A & = & \displaystyle \frac{\pi^2}{4}
& B_t^A & = & \displaystyle \frac{4 \pi^2}{m_t^2}
\end{array}
\end{equation}
These constants correspond to the lowest order coefficients of the expansions
(\ref{eq:anbn}) in $\alpha_s$.
As noted above the large value of the $t$--quark ($W$--boson) width
$\Gamma_t$ ($\Gamma_W$) modifies the properties of the processes
$A\to \gamma\gamma$ and $H\to \gamma\gamma$ in the threshold region.
In the non--relativistic approximation the denominators of the quark (boson)
propogators with momentum $p=(m+\epsilon ,\vec p)$ take the form
$2m(\epsilon -\frac{\vec p^2}{2m}+i\frac{\Gamma}{2})$ (see \cite{FaKh}).
The calculation of the amplitude $F$ is analogous to the case of zero
width (see Appendix) leading to the simple replacement
\begin{equation}
E \rightarrow E+i \Gamma ,
\end{equation}
in particular
$p\to p=\sqrt{m(-E-i\Gamma)}=p_- +i p_+$
with $p_\pm = \sqrt{\frac{m}{2} ( \sqrt{E^2 + \Gamma^2} \pm E ) }$.
 From eq.(\ref{eq:meix}) we obtain for the pseudoscalar case \cite{FaKh}:
\begin{eqnarray}
\Re e~G_t(0,0;E+i\Gamma_t) & = &
-\frac{m_t p_-}{4\pi} + \frac{m_t p_0}{4\pi} \log\left(
\frac{p_0^2}{p_+^2 + p_-^2} D^2 \right) \non \\ \non \\
& & + \frac{m_t p_0^2}{2\pi}
\sum_{n=1}^\infty \frac{p_- - p_n}{n^2 ((p_- - p_n)^2 + p_+^2 )}
\non \\ \non \\
\Im m~G_t(0,0;E+i\Gamma_t) & = &
\frac{m_t p_+}{4\pi} + \frac{m_t p_0}{2\pi} \arctan \frac{p_+}{p_-}
\non \\ \non \\
& & + \frac{m_t p_0^2}{2\pi} \sum_{n=1}^\infty
\frac{p_+}{n^2 ((p_- - p_n)^2 + p_+^2 )} \label{eq:reim}
\end{eqnarray}
\begin{displaymath}
\mbox{with~} \hspace{0.5cm} p_n = \frac{p_0}{n} \hspace{1cm}
\mbox{and~} \hspace{0.5cm} p_0 = \frac{2}{3} m_t \alpha_s
\end{displaymath}
For our numerical calculation we have used the Standard Model
$t$ quark width \cite{bizer}:
\begin{equation}
\Gamma_t = \frac{G_F m_t^3}{8\pi \sqrt{2}} \left(\frac{2k}{m_t} \right)
\left[ (1-x^2)^2 + (1+x^2)y^2 -2y^4 \right]
\end{equation}
where $x=m_b/m_t$, $y=m_W/m_t$ and $k$ is the 3--momentum of the
decay products in the $t$ rest frame: $k=\frac{1}{2}\sqrt{[1-(x+y)^2]
[1-(x-y)^2]}$. Further we choose
$m_t=140$ Ge$\!$V, $m_b=5$ Ge$\!$V, $m_W=80.14$ Ge$\!$V,
$\alpha_s=0.15$ and $\Gamma_W =2.15$ Ge$\!$V.

The results for $\Re e F^H_W$, $\Im m F_W^H$ and $\Re e F_t^A$, $\Im m F^A_t$
are plotted in figs.2 and 3 respectively.
The effect of the $W$ decay width (fig.2) to the scalar decays
$h,H\to \gamma\gamma$
amounts to about 15\% at the $W^+W^-$--threshold rapidly decreasing apart from
the threshold region.

We observe from fig.3 that $\Im m F^A_t$ contains a peak at the energy of
the lowest $(t \bar t)$--bound state,
but $\Re e F_t^A$ shows a maximum on the left--hand side of this energy level
and a minimum on the right--hand side.
This behaviour can easily be understood in the narrow--width
approximation ($\Gamma_t \ll |E|$) leading to a qualitative
picture of the result. Consider $F^A_t$ at $E\to E_n$, where $E_n$
is the bound state energy of level $n$. The Green's function contains a pole at
$E=E_n-i\Gamma_t$. Keeping the leading term of $G(0,0;E)$ only
we obtain
\begin{equation}
F_t^A(E) = A_0 + B_0 \frac{|\Psi_0 (0)|^2}{E_0-E-i\Gamma_t} \label{eq:fapsi}
\end{equation}
$\Psi_0(0)$ denotes the lowest level wave function of the Coulomb potential
at the origin, for
arbitrary $n$ given by $|\Psi_n(0)|^2=k_n^3/\pi$ with
$k_n=\frac{2}{3}m_t\alpha_s/(n+1)$.
The first term in (\ref{eq:fapsi}) is a real constant, the imaginary part of
the second develops
a maximum at $E=E_0$, but the real part vanishes there showing a maximum
for $E<E_0$ and a minimum at $E>E_0$.

\section{QCD Corrections}
%        ===============

In this section the QCD corrections to the $A\gamma\gamma$ amplitude
at the threshold $m_A=2m_t$ are calculated.
First we consider the exact result for the two--loop QCD corrections
to the $A\gamma \gamma$ vertex \cite{DSZ}. In order to obtain the radiative
corrections to the coefficient $A$ of eq.(\ref{eq:green}) we must extract
the contribution of one Coulomb gluon exchange from the exact two--loop result
[$B\frac{m_tp_0}{2\pi}\log(\frac{m_tD}{p})$], which is already absorbed in the
Green's function (\ref{eq:meix}). After this subtraction we obtain
\begin{eqnarray}
A & = & A_0 \left(1 + a \frac{\alpha_s}{\pi} \right) \non \\ \non \\
a & \approx & -4.86 \hspace{1cm} \mbox{~for~} \hspace{0.1cm} D=1
\end{eqnarray}
We should note that the constant $a$ depends on the parameter $D$,
but the total result for the form factor $F_t^A$ does not depend on $D$.
There are three sources of radiative corrections to the second term in
eq.(\ref{eq:green}):
\begin{enumerate}
\item QCD radiative corrections to the static heavy quark--antiquark potential
$V(r)$ \cite{Fisc,Bill},
\item QCD radiative corrections to the Born width of the top quark \cite{Jeza},
\item hard QCD radiative corrections to the $H\to t\bar t$ and
$t\bar t\to \gamma \gamma$ amplitudes \cite{Feyn,Braa}.
\end{enumerate}
In the following we analyze these aspects in detail.
The QCD radiative corrections to the quark--antiquark potential are the
dominant contributions. First we note that the non--relativistic
potential is of perturbative nature, because non--perturbative effects
are suppressed by the large top decay width in the $t\bar t$ system
\cite{FaYa,Pesk}.
In leading order of $\alpha_s$ the potential is given by the Coulomb form
\begin{equation}
V(r)= -\frac{4}{3}\frac{\alpha_s}{r},
\end{equation}
In next--to--leading order $V(r)$ depends on the renormalization scale
and obeys the corresponding renormalization group equation.
We use the QCD potential with
running coupling $\alpha_s$:
\begin{equation}
\alpha_s(r)=\frac{4\pi}{b_0\log((\Lambda r)^{-2})+b_1/b_0
\log\log\left((\Lambda r)^{-2}\right)}
\end{equation}
where $b_0=11-\frac{2}{3}n_f$, $b_1= 102-\frac{38}{3}n_f$ are the first
coefficients in the perturbative expansion of the QCD $\beta$--function.
The QCD scale is given by \cite{Pesk}
\begin{eqnarray}
\Lambda & = & \Lambda_{\overline{MS}}\exp\left(\frac{C}{2}\right)
\approx 2.43~\Lambda_{\overline{MS}} \hspace{0.5cm} (n_f=5) \non \\ \non \\
\mbox{with~} \hspace{1cm} C & = & \frac{1}{b_0}\left(\frac{31}{3}-
\frac{10}{9} n_f\right)+2\gamma_E \non \\ \non \\
\mbox{and~}\hspace{1cm} \Lambda_{\overline{MS}} & \approx &
\left(\frac{4\pi}{b_0\alpha_s(M_Z)}\right)^{[b_1/(2b_0^2)]}
M_Z \exp\left[-\frac{2\pi}{b_0\alpha_s(M_Z)}\right]
\end{eqnarray}
Following Ref.\cite{Pesk} we perform a simple
regularization of the pole at $\Lambda r\approx 0.785$ by replacing
$\Lambda r\to a \tanh (\Lambda r/a)$ with $a=0.3$.
The final form of the QCD--potential can be expressed as
\begin{equation}
V(r)= -\frac{4}{3}\frac{\alpha_s(r)}{r},
\end{equation}
The Green's function $G_t(\vec r,\vec r\,';E)$ was calculated numerically
with an accuracy
of $10^{-5}$, where we checked the results for the real and imaginary parts
using a Coulomb potential with constant $\alpha_s$ against the analytical
results
(\ref{eq:reim}). As a further check
we compared our result for $\Im m G_t(0,0;E)$ using a potential with
running $\alpha_s$ with the results of \cite{Pesk,Jeza1,Sumi}
and found full agreement.
Only the energy dependent part of $\Re e G_t(0,0;E)$ can be determined by
our numerical program, whereas the unknown constant not depending on $E$ can be
fixed by comparing our numerical result with the analytical expressions
(\ref{eq:reim})
at large energy $E$, because both results must match in ${\cal O}(\alpha_s)$.
The numerical results for $\Re e F^A_t$ and $\Im m F^A_t$ are plotted in fig.4
choosing $\alpha_s(m_Z)=0.12$.

The QCD radiative corrections to the Born width of the top quark were
calculated in Ref.\cite{Jeza} and are negligible in our analysis.
The third contribution leads to a correction to the constant $B$ in
eq.(\ref{eq:green})
\begin{equation}
B = B_0 \left(1+b\frac{\alpha_s}{\pi}\right)
\end{equation}
The coefficient b can be obtained analytically from the well--known results
of Refs.\cite{Feyn} and \cite{Braa}, because the real corrections to their
results belong to a $P$--wave contribution and therefore vanish at threshold.
The hard corrections at threshold to the process $t\bar t\to \gamma \gamma$
are given by
\cite{Feyn}
\begin{equation}
 1-\frac{\alpha_s(2m_t)}{\pi}\left[\frac{C_F}{2}\left(5-\frac{\pi^2}{4}\right)
\right]
\end{equation}
and the corresponding ones to $A\to t\bar t$ by \cite{Braa}
\begin{equation}
 1-\frac{\alpha_s(2m_t)}{\pi}\left(3~\frac{C_F}{2}\right)
\end{equation}
The coefficient $b$ is determined by the sum of both contributions:
\begin{equation}
 b= -\frac{C_F}{2}\left(8-\frac{\pi^2}{4}\right)
\end{equation}
and has been compared to the result of the exact two--loop calculation
\cite{Djou} finding agreement of the two.

\section{Conclusions}
%        ===========

We have analyzed threshold effects to the photonic decay modes $\phi\to
\gamma\gamma$ of the scalar and pseudoscalar Higgs particles in the Standard
Model [$\cal SM$] and the minimal supersymmetric extension [$\cal MSSM$].
These contributions are generated by the finite decay widths of the top
quark and $W$ boson as well as strong Coulomb interactions among $t\bar t$
pairs giving rise to bound state formation in the threshold region.
[Effects due to the mixing between the Higgs fields and the bound states are
important \cite{DreHi}, but have not been considered in our analysis.] The
top--loop does not contribute significantly in a $P$--wave to the scalar
Higgs decays, resulting
in negligible threshold effects. $W^+W^-$ pair production in the scalar case
proceeds via an $S$--wave, so that the inclusion of the $W$ decay width leads
to threshold effects of ${\cal O}(10\%)$ near the $W^+W^-$--threshold.

The pseudoscalar decay $A\to\gamma\gamma$ is generated by a virtual $t\bar t$
pair in an $S$--wave. Therefore the inclusion of the top decay width and strong
Coulomb interactions leads to large and important contributions to the
amplitude. We have presented the full form factor performing a
non--ralativistic approximation in the threshold region. The calculated effects
are large and cannot be neglected, if the pseudoscalar mass $m_A$ is close to
twice the top quark mass $m_t$ [$|m_A-2m_t| \ll m_A$]. This analysis
completes previous calculations of perturbative QCD corrections to the
photonic Higgs decays at the two--loop level \citer{Zhen,DSZ}.

\section{Appendix}
%        ========

In this appendix we present the method for our calculation of the amplitudes
for $H \to \gamma \gamma$ and $A \to \gamma \gamma$ near the $W^+W^-$ or
$t\bar t$ threshold.
First we consider the process $A \to \gamma \gamma$ near threshold
[$m_A=2m_t+E$ with $|E| \ll m_A$].
$F^A_t(E)$ will be calculated in the Coulomb gauge, which is the most
convenient one for non--relativistic problems.
The general point of our approach is the division of all Feynman diagrams
into two different classes [fig.5]:
\begin{description}
\item[(i)] Class A contains all diagrams without Coulomb gluon exchanges.
\item[(ii)] Class B contains only those which are built--up by
at least one Coulomb gluon exchange.
\end{description}
For the calculation of diagrams in class A we may use usual perturbative
techniques and expand the result in the small ratio $x=E/m_t$, because it is
a regular function of $x$ at the origin $x=0$.
It is well--known that diagrams in class B develop singularities at $x=0$,
which correspond to bound states below threshold
and strong interaction effects above threshold, resulting in a
logarithmic divergence at $x=0$.
In order to calculate diagrams of Class B we divide the integration over
the top momentum $p_t$ into two regions:
\begin{description}
\item[(B1)] $|p_t| \gg Q$
\item[(B2)] $|p_t| \ll Q$
\end{description}
$Q$ is an arbitrary scale obeying the condition
$p_{non-rel}\ll Q\ll p_{rel}$ with $p_{rel}\sim m_t$ [$p_{non-rel}\sim
\sqrt{Em_t}$] determining the characteristic scale of relativistic
[non--relativistic] phenomena. Region B1 generates no singularity in $E$ and
we can neglect all its non--relativistic effects.
In region B2 we may use the non--relativistic approximation
for the $p_t$--integration, in the following analyzed in detail.
The form factor $F_t^A$ can be written as [fig.6]:
\begin{equation}
F_t^{A}=b\int \frac{d^4p_t}{(2\pi)^4}Tr \left\{S_t(p_t)\Gamma_{At\bar t}
S_t(p_{\bar t})\Gamma_{t\bar t \gamma \gamma}\right\} \label{eq:form}
\end{equation}
where $S_t(p)$ is the fermion propagator and $\Gamma_{At\bar t}$
($\Gamma_{t\bar t \gamma \gamma}$) the vertex operator of the corresponding
transition. We choose the fermion propagator in its non--relativistic
form
\begin{equation}
S_t(p)=\frac{1+\gamma_0}{2}\frac{i}{\varepsilon-\frac{\vec p^2}{m_t}+
i\frac{\Gamma_t}{2}} \hspace{1cm} \mbox{with} \hspace{1cm}
p=(m_t+\varepsilon,\vec p) \label{eq:prop}
\end{equation}
We may adopt $\Gamma_{At\bar t}$ in the Born approximation, but must
take into account all Coulomb gluon exchanges in the
$\Gamma_{t\bar t \gamma \gamma}$--vertex.
One obtains the followig result from this procedure:
\begin{equation}
\Gamma_{t\bar t\gamma\gamma}(\vec p,E)=\Gamma^0_{t\bar t \gamma\gamma}~
\left\{ \frac{\vec p^2}{m_t}-E-i\Gamma_t\right\}~G_t(\vec p ;E) \label{eq:vert}
\end{equation}
where $\Gamma^0_{t\bar t \gamma\gamma}$ is the lowest order coupling of two
photons to a $t\bar t$ pair at threshold, being independent of $\vec p$ and
$E$, and $G_t(\vec p;E)$ the Fourier transform of the
$S$--wave Green's function of the non--relativistic
Schr\"odinger equation with the Coulomb potential
\begin{eqnarray}
(\hat H - E -i\Gamma_t ) G_t(\vec r,\vec r\,';E) & = & \delta (\vec r-\vec
r\,')
\non \\ \non \\
\mbox{with} \hspace{0.5cm} \hat H & = & -\frac{\nabla^2}{m_t} + V(r)
\non \\ \non \\
V(r) & = & -\frac{4}{3}\frac{\alpha_s}{r}
\non \\ \non \\
G_t(\vec p;E) & = & \int d^3\vec r~G_t(\vec r,\vec r\,'=0;E)~
e^{-i\vec p \vec r}
\end{eqnarray}
After substituting eqs.(\ref{eq:prop}) and (\ref{eq:vert}) into
eq.(\ref{eq:form}) the
$p_t^0$--integration can be performed explicitly by taking the residue
of the pole at $p_t^0=m_t+\frac{\vec p^2}{m_t}-i\frac{\Gamma_t}{2}$.
Adding the contributions of class B and those of the class A diagrams
we obtain as the final result
\begin{equation}
F^A_t(E) = A + B~G_t(0,0;E),
\end{equation}
where $A$ and $B$ are real constants, which can be expanded in a
perturbative series:
\begin{equation}
A=\sum_{n=0}^\infty A_n \left(\frac{\alpha_s}{\pi}\right)^n \hspace{2cm}
B=\sum_{n=0}^\infty B_n \left(\frac{\alpha_s}{\pi}\right)^n
\end{equation}
The coefficients $A_n$ and $B_n$ can be determined from the comparison
with the usual perturbative QCD corrections.
The calculation of the amplitude $F_W^H$ is performed in an analogous way
without the contribution of the Coulomb potential $V(r)$
by taking into account the $W$ decay width $\Gamma_W$ only.

\noindent
{\Large \sc Acknowledgements}
%           ================

\noindent
The authors are grateful to P.M.Zerwas for fruitful discussions and careful
reading of the manuscript. K.M.~and O.Y.~would like to thank P.M.Zerwas for
the warm hospitality at DESY, Hamburg.

\noindent
{\Large \sc Figure Captions}
%           ===============

\begin{description}
\item[Fig.1:] Generic Feynman diagrams of the $H\gamma\gamma$ and
$A\gamma\gamma$ couplings in lowest order: (a) top--loop contributing
to the scalar $H\gamma\gamma$ and the pseudoscalar $A\gamma\gamma$
coupling, (b) $W$--loop contributing to the scalar $H\gamma\gamma$ coupling.

\item[Fig.2:] $W$--loop form factor $F_W^H$ of the scalar Higgs decays $h,H
\to \gamma\gamma$ plotted versus the distance $E=m_H-2m_W$ to the
$W^+W^-$--threshold. The results including the decay width $\Gamma_W$ are
compared to the lowest order perturbative ones.

\item[Fig.3:] Top--loop form factor $F_t^A$ of the pseudoscalar Higgs decay $A
\to \gamma\gamma$ plotted against the deviation $E=m_A-2m_t$ from the
$t\bar t$--threshold for a QCD--potential with constant $\alpha_s=0.15$.
The results including the complete Green's function $G_t(0,0;E+i\Gamma_t)$
(HO) are compared to the lowest order perturbative ones (LO).

\item[Fig.4:] Same as in Fig.2, but with a QCD--potential using a running
$\alpha_s$ normalized to $\alpha_s(M_Z)=0.12$.

\item[Fig.5:] Division of the gluon exchange diagrams contributing to the
$A\gamma\gamma$ coupling into the classes A and B. $g_T$ denotes
transverse and $g_L$ longitudinal gluon exchange in the Coulomb gauge.

\item[Fig.6:] Diagrammatic representation of the $A\gamma\gamma$ coupling
in terms of the vertex operators $\Gamma_{t\bar t \gamma\gamma}$
and $\Gamma_{A t\bar t}$.
\end{description}

\end{document}